\documentclass[aps,prd,twocolumn,preprintnumbers,floatfix]{revtex4-1}

\usepackage{graphicx}
\usepackage{dcolumn}
\usepackage{bm}
\usepackage[mathlines]{lineno}
\usepackage{epsfig}  
\usepackage{color}
\usepackage{slashed}
\usepackage{float}
\usepackage[table]{xcolor}
\usepackage{amsmath}
\usepackage{soul}
\usepackage{braket}
\allowdisplaybreaks

\usepackage{cancel}
\usepackage{hyperref}
\usepackage{orcidlink}
\hypersetup{colorlinks=true, linkcolor=blue, citecolor=blue, urlcolor=magenta}

\newcommand{\BzBzbar}{\( B^{0}\text{-}\bar{B}^{0} \)}

\newcommand{\ba}{\begin{array}}
\newcommand{\ea}{\end{array}}
\newcommand{\bea}{\begin{eqnarray}}
\newcommand{\eea}{\end{eqnarray}}

\newcommand{\be}{\begin{equation}}
\newcommand{\ee}{\end{equation}}

\begin{document}

\title{Entanglement metrics for a $B$-meson system}

\author{Aashish Joshi\orcidlink{0009000598687887}}
\email{p24ph0001@iitj.ac.in}
\affiliation{Indian Institute of Technology Jodhpur, Jodhpur 342037, India}

\author{Prisha\orcidlink{0009000955217761}}
\email{p23ph0008@iitj.ac.in}
\affiliation{Indian Institute of Technology Jodhpur, Jodhpur 342037, India}

\author{Neetu Raj Singh Chundawat\orcidlink{0000000300926260}}
\email{chundawat@ihep.ac.cn}
\affiliation{Institute of High Energy Physics, Chinese Academy of Sciences, Beijing 100049, China}
\affiliation{Kaiping Neutrino Research Center, Guangdong 529386, China}

\author{Jitendra Kumar\orcidlink{000000028465433X}}
\email{jkumar@iitj.ac.in}
\affiliation{Indian Institute of Technology Jodhpur, Jodhpur 342037, India}

\begin{abstract}
The $B$-factory experiments operate at electron-positron colliders with beam energies precisely tuned for optimal \BzBzbar\ meson pair production. These \BzBzbar\ meson pairs are entangled and offer a unique opportunity to explore quantum correlations and examine the foundational aspects of quantum mechanics. In this work, we present a comprehensive analysis of entanglement metrics in the $B$-meson system within the framework of open quantum systems, which introduces decoherence due to interactions with the environment. We examine several distinct entanglement measures, each highlighting different facets of quantum entanglement and its sensitivity to decoherence effects. We further analyze the impact of decoherence by systematically varying the decoherence parameter across different scales.

\end{abstract}

\maketitle

\section{Introduction}
\label{intro}
Quantum entanglement is one of the most striking non-classical features of quantum theory, where the state of one subsystem is intrinsically correlated with that of another, regardless of the spatial separation between them. This phenomenon lies at the heart of the Einstein-Podolsky-Rosen (EPR) paradox and has profound implications for the nature of reality and causality \cite{Einstein:1935rr,Bell:1964kc}. Bell’s inequalities are widely used as proxies for detecting non-local behavior and serve as empirical evidence that quantum mechanics cannot be explained by local hidden-variable theories. Numerous experimental tests performed over the last few decades have consistently shown violations of Bell inequalities, challenging classical notions of locality and establishing the inherently non-local nature of quantum correlations \cite{Clauser:1974tg,Clauser:1978ng,Aspect:1981zz,Aspect:1981nv,Aspect:1982fx,Tittel:1998ja}. More recently, landmark loophole-free Bell-test experiments have simultaneously closed both the detection-efficiency and locality loopholes, providing the most compelling modern confirmation of quantum non-locality \cite{Hensen:2015ccp,Giustina:2015yza,Stevens:2015awv}. Together, these foundational studies demonstrate that entanglement is not merely a theoretical curiosity, but a robust physical phenomenon, motivating its exploration in environments beyond traditional optical systems, including high-energy physics and in relativistic settings.

This effort to explore quantum correlation has expanded in the domain of high-energy physics~\cite{Gabrielli:2025ybg,Fabbrichesi:2025aqp,Cheng:2025zaw,Maltoni:2024csn,Gabrielli:2026tnl,Afik:2025ejh,Fang:2024ple,Barr:2025avs,Cheng:2024rxi,Cheng:2025zcf,Gu:2025ijz,Wu:2025dds,Fabbrichesi:2023idl,Gabrielli:2024kbz,Aoude:2025jzc,Fabbrichesi:2025zpw,Fabbrichesi:2025psr,Afik:2025grr,Banerjee:2014vga}, where massive and unstable particles offer a fundamentally different testbed system in comparison to traditional optical setups involving entangled photons and, in some cases, electrons in solid-state systems. A recent review highlights the growing role of entanglement studies in particle physics, including tests of Bell-inequality violation and detection of entanglement at colliders~\cite{Barr:2024djo}. These studies demonstrate that entanglement can be systematically probed across different hadronic systems.

Recently, ATLAS and CMS experiments at CERN measured spin correlations in top-antitop (\(t\bar{t}\)) pairs, where entanglement can be inferred through decay product angular distributions \cite{ATLAS:2023fsd, CMS:2024zkc, CMS:2024pts}, building on phenomenological studies proposing tests of quantum correlations and entanglement in top-quark pairs~\cite{Afik:2020onf,Fabbrichesi:2021npl,Severi:2021cnj,Aguilar-Saavedra:2022uye,Rai:2025qke}. The entangled \(\tau^+\tau^-\) pairs produced at Belle II and BEPC also provide a promising system to study quantum correlations involving leptons \cite{Ehataht:2023zzt,Han:2025ewp}. Neutrino oscillations, on the other hand, offer a platform where quantum correlations are explored in the neutrino sector \cite{Akhmedov:2010ua,Blasone:2007vw,Ettefaghi:2020otb, Alok:2014gya, Alok:2025qqr}. Furthermore, EPR-entangled neutral meson pairs such as \( K^0\bar{K}^0 \) from \( \phi \) decays and \BzBzbar\ from \( \Upsilon(4S) \) decays are especially compelling for two reasons. First, their nonzero mass distinguishes them from massless photon systems. Second, their strong hadronic interactions lead to comparatively high tagging and reconstruction efficiencies, which decrease, though do not yet completely close, the detection loophole that has historically affected photon-based Bell tests \cite{Bramon:2001tb,Bramon:2005mg}. 

The entangled \BzBzbar\ system produced at $B$-factory experiments such as Belle at KEK, Japan, BaBar at SLAC, USA, and the ongoing Belle II at SuperKEKB, Japan, acts as a natural testbed and is special in the sense that it exists only briefly before decaying, and entanglement occurs in the flavor (particle-antiparticle) basis~\cite{Datta:1986ut,Pompili:1999tv,Go:2003tx,Bertlmann:2004cr,Bramon:2004pp}. In determining observables related to neutral $B$ mesons and in the aforementioned proposals, the assumption of perfect entanglement is often made. However, any real system will inevitably interact with its environment, leading to decoherence, effectively suppressing quantum interference and entanglement. The decoherence, therefore, would potentially alter time-dependent correlations in their decay. At the fundamental level, quantum decoherence can arise from space-time foam effects, both in critical string theories and in quantum loop gravity \cite{Ellis:1997jw,Mavromatos:2004sz,Gambini:2003pv}. Additionally, previous studies suggest that quantum decoherence may also be linked to dark energy \cite{Mavromatos:2003hr}. Decoherence can be probed in systems exhibiting quantum interference, such as flavor oscillations in neutral mesons and neutrinos. These oscillating systems are sensitive to coherence loss, unlike typical photon systems, which lack such dynamical structure. 

An important question to explore is whether the entanglement of the \BzBzbar\ pair degrades over its lifetime \cite{Bertlmann:2004yg}. A closely related issue is how to quantify the embedded entanglement itself~\cite{Verstraete:2002mhg, Eltschka:2014via, Plenio:2007zz, Laurell:2024oig, Zhao:2024ddp, Christandl:2003med,Tucci:1999scb,Tucci:2002gzf,Muller-Lennert:2013liu} and unfortunately, there exists no single, straightforward measure that captures all facets of entanglement, reflecting the complexity and richness of quantum correlations. Entropy-based measures such as entanglement entropy, Rényi entropy, and relative entropy of entanglement are widely used, particularly for bipartite pure states, while operational measures like entanglement of formation and entanglement cost relate to resource requirements under local operations and classical communication (LOCC). Distance-based measures, including the geometric measure and fidelity-based approaches, assess how far a state lies from separability. For mixed states, negativity and logarithmic negativity are used due to their computability and purity, while not a direct entanglement measure, provides insight into the mixedness of subsystems. Structural approaches such as Schmidt coefficients and majorization criteria help in capturing the internal structure and ordering of entanglement. Finally, in multipartite or open systems, global entanglement measures and entanglement witnesses play an important role as they can characterize complex or environment-influenced entanglement without requiring full state reconstruction.

Quantum correlations are generally fragile against environmental decoherence~\cite{Carvalho:2004ozj}, which implies that the entanglement shared between the subsystems is expected to degrade when the system is treated as an open quantum system. In this work, we study the time evolution of the entangled neutral $B$-meson pair within the open quantum system formalism, incorporating decoherence effects~\cite{Caban:2005ue,Alok:2015iua,Naikoo:2018vug,Alok:2024amd,Caban:2007je,Caban:2006ij,Alok:2025duw,Panda:2025obl}. Earlier studies have investigated entanglement in the $B$-meson system and decoherence effects in oscillation and decay observables. Our study instead focuses on a systematic comparison of the sensitivity of different entanglement metrics to decoherence in this system, showing that some measures remain insensitive to the decoherence parameter $\lambda$ while others decay monotonically with it, a distinction directly relevant for selecting observables in future experimental searches for decoherence. The time evolution of the density matrix of the entangled \BzBzbar\ meson pair is described using the Kraus operator formalism, with these Kraus operators containing a decoherence parameter along with other parameters specific to neutral-meson mixing and decay. The open-quantum-system framework adopted here does not reflect any specific origin of decoherence; instead, $\lambda$ is treated as an effective, model-independent phenomenological parameter that captures the net effect of system-environment interactions on the \BzBzbar\ system, irrespective of their fundamental origin.

The plan of this work is as follows. Section~\ref{sec: BBevol} introduces the time evolution of the entangled $B$-meson system within the framework of open quantum systems. In Section~\ref{sec:measures}, we provide a detailed account of each entanglement measure used in our analysis and compute them for the system under consideration. Finally, Section \ref{sec:conc} presents our concluding remarks.

\section{time evolution of  the \BzBzbar\ system}
\label{sec: BBevol}
In this section, we explore the time-dependent evolution of the \BzBzbar\ meson system within the framework of open quantum systems. We begin by assuming that the system is initially prepared in a singlet state, representing a maximally entangled configuration. The initial density matrix of the system composed of two neutral mesons labeled as $A$ and $B$, is given by,
\begin{equation}
\rho_{AB}(0) = |\psi\rangle \langle \psi|,
\end{equation}
 where the singlet state \(|\psi\rangle \) is defined as,
\begin{equation}
|\psi\rangle = \frac{1}{\sqrt{2}} \left( |B^0\rangle \otimes |\bar{B}^0\rangle - |\bar{B}^0\rangle \otimes |B^0\rangle \right).
\end{equation}

In this notation, the first particle in each ket moves to the left and the second to the right. Here, \(|B^0\rangle\) and \(|\bar{B}^0\rangle\) represent the flavor eigenstates of the neutral \( B \) mesons. The Hilbert space of the two neutral \( B \) mesons can be expressed as,
\begin{eqnarray}
\mathcal{H} 
&=& (\mathcal{H}_A \oplus \mathcal{H}_0) \otimes (\mathcal{H}_B \oplus \mathcal{H}_0)\nonumber\\
&=& (\mathcal{H}_A \otimes \mathcal{H}_B) \oplus (\mathcal{H}_A \otimes \mathcal{H}_0 \oplus \mathcal{H}_0 \otimes \mathcal{H}_B) \oplus (\mathcal{H}_0 \otimes \mathcal{H}_0). \nonumber
\end{eqnarray}
This decomposition reveals that the system can exist in a two-particle state \((\mathcal{H}_A \otimes \mathcal{H}_B)\), a one-particle state \((\mathcal{H}_A \otimes \mathcal{H}_0 \oplus \mathcal{H}_0 \otimes \mathcal{H}_B)\), or the vacuum state \((\mathcal{H}_0 \otimes \mathcal{H}_0)\). The dimensionalities of these subspaces are \(\dim(\mathcal{H}_A \otimes \mathcal{H}_B) = \dim(\mathcal{H}_A \otimes \mathcal{H}_0 \oplus \mathcal{H}_0 \otimes \mathcal{H}_B) = 4\) and \(\dim(\mathcal{H}_0 \otimes \mathcal{H}_0) = 1\).

Given that the \BzBzbar\ system is open, its time evolution is described by a completely positive map in the operator-sum representation,
\begin{equation}
\rho_{AB}(t) = \sum_{i,j} E_{ij}(t)\,\rho_{AB}(0)\,E_{ij}^\dagger(t),
\end{equation}
where $E_{ij}(t)$ are Kraus operators. For non-interacting particles, the dynamics factorize as
\begin{equation}
E_{ij}(t) = E_i(t) \otimes E_j(t).
\end{equation}

The operators $E_i(t)$ encode the non-unitary effects of decay and flavor oscillation, ensuring complete positivity and a trace non-increasing evolution consistent with the instability of mesons. In practical applications, the density matrix is restricted to the surviving two-particle sector $\mathcal{H}_A \otimes \mathcal{H}_B$, since only this component contributes to measurable entanglement.

In this sector, the Kraus‐map normalization condition,
\begin{equation}
\sum_{i,j}E_{ij}^\dagger(t)E_{ij}(t) = I\otimes I,
\end{equation}
follows directly from the single-particle completeness relation $\sum_i E_i^\dagger(t)E_i(t) = I$. This representation \cite{Alok:2024amd,Caban:2007je,Caban:2006ij} captures the evolution of an open quantum system, which is not unitary in contrast to the Hamiltonian evolution of closed systems. Real physical systems are always entangled with their surrounding environment, often referred to as the reservoir, which motivates the need for this framework when describing neutral-meson dynamics.

Consider a large system \(S\) comprising two subsystems \(S_A\) and \(S_B\). At a given time \(t\), let the quantum states corresponding to \(S\), \(S_A\), and \(S_B\) be represented by \(\rho(t)\), \(\rho_A(t)\), and \(\rho_B(t)\), respectively. Then,
\begin{equation}
\rho_A(t) = \text{Tr}_B\{\rho(t)\} \quad \text{and} \quad \rho_B(t) = \text{Tr}_A\{\rho(t)\}.
\end{equation}
Since the total system is unitary, its evolution is given by
\begin{equation}
\rho(t) = U(t)\rho(0)U^\dagger(t),
\end{equation}
where \(U(t)\) is a unitary operator. The evolution of system \(S_A\) will look like
\begin{equation}
\rho_A(t) = \text{Tr}_B\{U(t)\rho(0)U^\dagger(t)\}.
\end{equation}

\noindent If it is possible to recast the above equation in the following form

\begin{equation}
\rho_A(t) = \sum_i E_i(t)\rho_A(0)E_i^\dagger(t),
\end{equation}

then the evolution of \(\rho_A(t)\) has a Kraus representation and is completely positive.

 The Kraus operators, initially developed for the $K$ meson system in ref. \cite{Caban:2007je}, have been utilized in our investigation to explore the $B$ meson system. The form of these Kraus operators is obtained to be
 \begin{widetext}
\begin{align*}
E_0 &= |0\rangle \langle0|, \\
E_1 &= \mathcal{C}_{1+} \left( |B^0\rangle \langle B^0| + |\bar{B}^0\rangle \langle \bar{B}^0| \right) + \mathcal{C}_{1-} \left( \frac{p}{q} |B^0\rangle \langle \bar{B}^0| + \frac{q}{p} |\bar{B}^0\rangle \langle B^0| \right), \\
E_2 &= \mathcal{C}_2 \left( \frac{p+q}{2p} |0\rangle \langle B^0| + \frac{p+q}{2q} |0\rangle \langle \bar{B}^0| \right), \\
E_3 &= \mathcal{C}_{3+} \left( \frac{p+q}{2p} |0\rangle \langle B^0| \right) + \mathcal{C}_{3-} \left( \frac{p+q}{2q} |0\rangle \langle \bar{B}^0| \right), \\
E_4 &= \mathcal{C}_4 \left( |B^0\rangle \langle B^0| + |\bar{B}^0\rangle \langle \bar{B}^0| + \frac{p}{q} |B^0\rangle \langle \bar{B}^0| + \frac{q}{p} |\bar{B}^0\rangle \langle B^0| \right), \\
E_5 &= \mathcal{C}_5 \left( |B^0\rangle \langle B^0| + |\bar{B}^0\rangle \langle \bar{B}^0| - \frac{p}{q} |B^0\rangle \langle \bar{B}^0| - \frac{q}{p} |\bar{B}^0\rangle \langle B^0| \right).
\end{align*}
The $\mathcal{C}$ coefficients are
\begin{align*}
\mathcal{C}_{1\pm} &=  \frac{1}{2} \left[ e^{-\left(2 i m_L + \Gamma_L + \lambda \right)t/2} + e^{-\left(2 i m_H + \Gamma_H + \lambda \right)t/2} \right], \\
\mathcal{C}_2 &= \sqrt{\frac{{\rm Re}[(p-q)/(p+q)]}{|p|^2-|q|^2} \left(1-e^{-\Gamma_L\,t}-\left(|p|^2-|q|^2\right)^2 \frac{\left|1-e^{-(\Gamma+\lambda-i\Delta m)t}\right|^2}{(1-e^{-\Gamma_H\,t})}\right)}, \\
\mathcal{C}_{3\pm} &= \sqrt{\frac{{\rm Re}[(p-q)/(p+q)]}{(|p|^2-|q|^2)(1-e^{-\Gamma_H\,t})}}\Big(1-e^{-\Gamma_H\,t} \pm \left(1-e^{-(\Gamma+\lambda-i\Delta m)t}\right)\left(|p|^2-|q|^2\right) \Big), \\
\mathcal{C}_4 &= \frac{e^{-\Gamma_L t/2}}{2} \sqrt{1 - e^{-\lambda t}}, \\
\mathcal{C}_5 &= \frac{e^{-\Gamma_H t/2}}{2} \sqrt{1 - e^{-\lambda t}}.
\end{align*}
\end{widetext}
Here, \(\Gamma\) is defined as the average decay width of the \(B^0\) meson system, calculated as \((\Gamma_L + \Gamma_H) / 2\). The parameter \(\Delta \Gamma\) represents the difference in decay widths between the light and heavy mass eigenstates of the \(B^0\) mesons, given by \(\Gamma_L - \Gamma_H\). Here, \(\Gamma_L\) and \(\Gamma_H\) correspond to the decay widths of the light state \(B^0_L\) and the heavy state \(B^0_H\), respectively. The light and heavy states are expressed as:
\begin{equation}
\begin{array}{l}
|B^0_L\rangle = p |B^0\rangle + q |\bar{B}^0\rangle, \\
|B^0_H\rangle = p |B^0\rangle - q |\bar{B}^0\rangle
\end{array}
\end{equation}
where \(B^0\) and \(\bar{B}^0\) are the flavor eigenstates of the \(B\) meson. The coefficients \(p\) and \(q\) are complex numbers that satisfy the normalization condition \( |p|^2 + |q|^2 = 1 \). The parameter \(\Delta m\) is the mass difference between the heavy and light states, defined as \(m_H - m_L\), where \(m_L\) and \(m_H\) are the masses of the \(B^0_L\) and \(B^0_H\) states, respectively. This mass difference is crucial for the oscillatory behavior observed in the \(B^0\) meson system.

The decoherence parameter $\lambda$ represents the interaction between the one-particle system and its environment within the open quantum description. In the Kraus-operator formalism described above, $\lambda$ enters as a damping parameter appearing in the exponential terms. Here, the limit $\lambda = 0$ corresponds to the perfectly coherent evolution of the meson. A nonzero value of this parameter can affect the entanglement and coherence properties of the $B^0$ meson system, and the value of $\lambda$ provides insight into the extent of environmental influence on the quantum state of the system. Together, the parameters $\Gamma_L$, $\Gamma_H$, $\Delta \Gamma$, $\Delta m$, and $\lambda$ describe the dynamic properties of the neutral $B$ meson system, including its decay, mass oscillation, and decoherence behavior.

It was shown that decoherence effects can modify the extraction of key observables such as the oscillation frequency $\Delta m$, the $CP$-violating parameter $\sin 2\beta$, and the decay-width difference $\Delta\Gamma$ in the $B$-meson system~\cite{Alok:2015iua}. Consequently, the same time-dependent oscillation and decay data used to measure these observables can also be used to constrain the decoherence parameter $\lambda$. In Ref.~\cite{Alok:2024amd}, several observables in semileptonic and non-leptonic $B$-meson decays were analyzed, which may also provide sensitivity to decoherence effects and hence can be used to determine or constrain the parameter $\lambda$. More broadly, the Kraus-operator formalism, originally developed for the $K^0$-$\bar{K}^0$ system~\cite{Caban:2007je,Caban:2006ij}, extends structurally to any mixing neutral meson pair, including $D^0$-$\bar{D}^0$ system.

For the density matrix formalism, the relevant object to calculate for entanglement quantification is the two-particle sector of the density matrix $\rho_{AB}(t)$ describing the state of the \BzBzbar\ system. That is, we must take into account only the part of the system that has not decayed up to time $t$. Accordingly, the full Hilbert space $\mathcal{H}$ is projected onto the surviving two-particle sector $\mathcal{H}_A \otimes \mathcal{H}_B$ using the projector $P_2$, defined as the projection operator onto $\mathcal{H}_A \otimes \mathcal{H}_B$.
\begin{widetext}
{ \begin{equation}
\rho_E(t) =  \frac{P_2 \rho_{AB}(t) P_2}{\mathrm{Tr}(P_2 \rho_{AB}(t))}\\
= \frac{(e^{2t\lambda} + 1) \left(1 -\left (|p|^2-|q|^2\right)^2\right)}{2(e^{2t\lambda} - \left (|p|^2-|q|^2\right)^2)} |\psi\rangle \langle \psi| + \frac{(e^{2t\lambda} - 1)\left(1 +\left (|p|^2-|q|^2\right)^2\right)}{2(e^{2t\lambda} -\left (|p|^2-|q|^2\right)^2)} |\chi\rangle \langle \chi|, 
\end{equation}}
\end{widetext}
where
\begin{equation}
P_2 = (|B^0\rangle \langle B^0| + |\bar{B}^0\rangle \langle \bar{B}^0|) \otimes (|B^0\rangle \langle B^0| + |\bar{B}^0\rangle \langle \bar{B}^0|), \nonumber
\end{equation}
and
{\small \begin{eqnarray}
|\chi\rangle &=& \frac{1 - |p|^2-|q|^2}{\sqrt{2(1 + (|p|^2-|q|^2)^2)}}  \left[ \left( \frac{p^2}{q^2} \right)  |B^0\rangle \otimes |B^0\rangle - |\bar{B}^0\rangle \otimes |\bar{B}^0\rangle \right].\nonumber
\end{eqnarray}}

Thus, the density matrix for the entangled \BzBzbar\ system is given by,
\begin{equation}
\rho_E(t) = A 
\begin{pmatrix}
|r|^4 a_-^{\prime} & 0 & 0 & -r^2 a_-^{\prime} \\
0 & a_+^{\prime} & -a_+^{\prime} & 0 \\
0 & -a_+^{\prime} & a_+^{\prime} & 0 \\
-r^2 a_-^{\prime} & 0 & 0 & a_-^{\prime}
\end{pmatrix},
\end{equation}
where
\begin{align}
a_{\pm}^{\prime} &= \left(e^{2\lambda t} \pm 1\right)\left(1 \pm \delta_L\right), \nonumber\\
A &= \frac{1 - \delta_L}{4\left(e^{2\lambda t} - \delta_L^2\right)},  \nonumber\\
\delta_L &= \frac{2\,\mathrm{Re}(\epsilon)}{1 + |\epsilon|^2},  \nonumber\\
r &= \frac{1 + \epsilon}{1 - \epsilon}. \nonumber
\end{align}

The density matrix $\rho_E(t)$ is written in the flavor basis 
$\{|BB\rangle, |B\bar B\rangle, |\bar B B\rangle, |\bar B\bar B\rangle\}$, 
where $|BB\rangle \equiv |B^0\rangle\otimes|B^0\rangle$ (and similarly for the other states), 
and it is constructed to be trace-preserving. The parameter $\epsilon$ represents a small 
CP-violating effect, typically of order $\sim 10^{-3}$ for $K$ mesons and around $10^{-5}$ for 
$B_{d,s}$ mesons. The parameter $\lambda$ captures the strength of decoherence. In the limit 
where CP violation is neglected, the density matrix simplifies considerably to:

\begin{equation}
\rho_E(t) = \frac{1}{4} 
\begin{pmatrix}
a_- & 0 & 0 & -a_- \\
0 & \quad a_+ & -a_+ & 0 \\
0 & -a_+ & \quad a_+ & 0 \\
-a_- & 0 & 0 & a_-
\end{pmatrix},
\label{denm-sim}
\end{equation}
where $a_{\pm} = 1 \pm e^{-2\lambda t}$.

The decoherence parameter $\lambda$ has not yet been constrained experimentally within the open-quantum-system framework adopted here. Belle has measured an analogous decoherence parameter using a different model, based on an ad hoc suppression of the quantum-mechanical interference term~\cite{Belle:2007ocp}. Within the Kraus-operator formalism employed in this work, an early phenomenological analysis obtained a rough estimate of order $\lambda_d \lesssim \mathcal{O}(0.1~\mathrm{ps}^{-1})$ for the $B_d^0$ system~\cite{Caban:2005ue}, which was subsequently obtained as $\lambda_d \lesssim 0.03~\mathrm{ps}^{-1}$ at 95\% C.L.~\cite{Alok:2015iua}. More recently, a dedicated fit to mixing- and $CP$-asymmetry data found $\lambda_d = 0.055 \pm 0.009~\mathrm{ps}^{-1}$ and $\lambda_s = 1.72 \pm 0.52~\mathrm{ps}^{-1}$ for the $B_d^0$ and $B_s^0$ systems, respectively~\cite{Alok:2025duw}. These estimates do not incorporate effects that could interfere with the extraction of the decoherence parameter, such as detector-level effects, systematic uncertainties, and resolution effects, which would be included in a dedicated experimental data fit.

The theoretical framework developed in this section, as well as all the entanglement measures computed in the following section, applies equally to the $B_d^0$ and $B_s^0$ systems. However, all numerical illustrations presented in this analysis are specific to the $B_d^0$--$\bar{B}_d^0$ system, for which the decoherence parameter $\lambda_d$ is small, and the time-dependent entanglement measures are expected to evolve slowly. We therefore present illustrations for the $B_d^0$--$\bar{B}_d^0$ system, showing the behavior of the $\lambda$-affected entanglement metrics for $\lambda_d = 0.01, 0.1$, and $1~\mathrm{ps}^{-1}$; the largest value substantially exceeds current phenomenological estimates and is retained only to expose the qualitative sensitivity of each measure over a wide decoherence range.

\section{Entanglement measures}
\label{sec:measures}
Entanglement admits several complementary quantifiers, since no single scalar measure captures all of its properties. In this section, we briefly summarize the main families of entanglement measures relevant to the entangled \BzBzbar\ system. We begin with entropy-based measures, including von Neumann and Rényi entropies and their relative variants, which provide a baseline characterization but are shown below to be insensitive to decoherence in the present case. We then consider mixedness and determinant-based indicators. Structural descriptors, distance-based measures, and multipartite-inspired quantities are subsequently reviewed. We conclude with negativity and logarithmic negativity, which provide a computable, time-sensitive characterization of entanglement degradation in the presence of decay.
\begin{figure*}[htb]
    \includegraphics[width=0.32\textwidth]{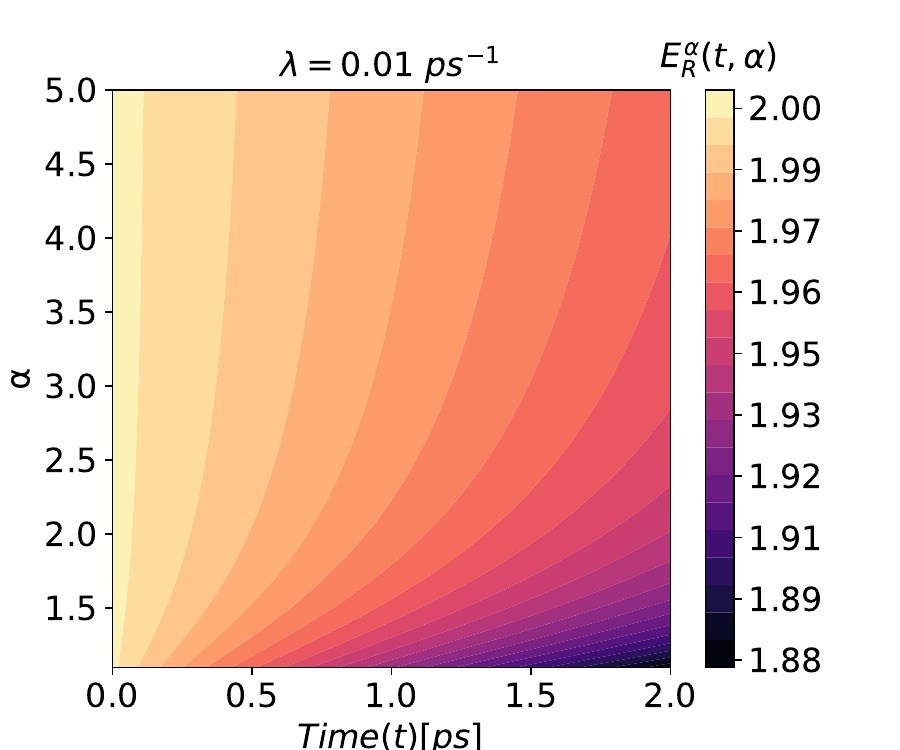}
    \includegraphics[width=0.32\textwidth]{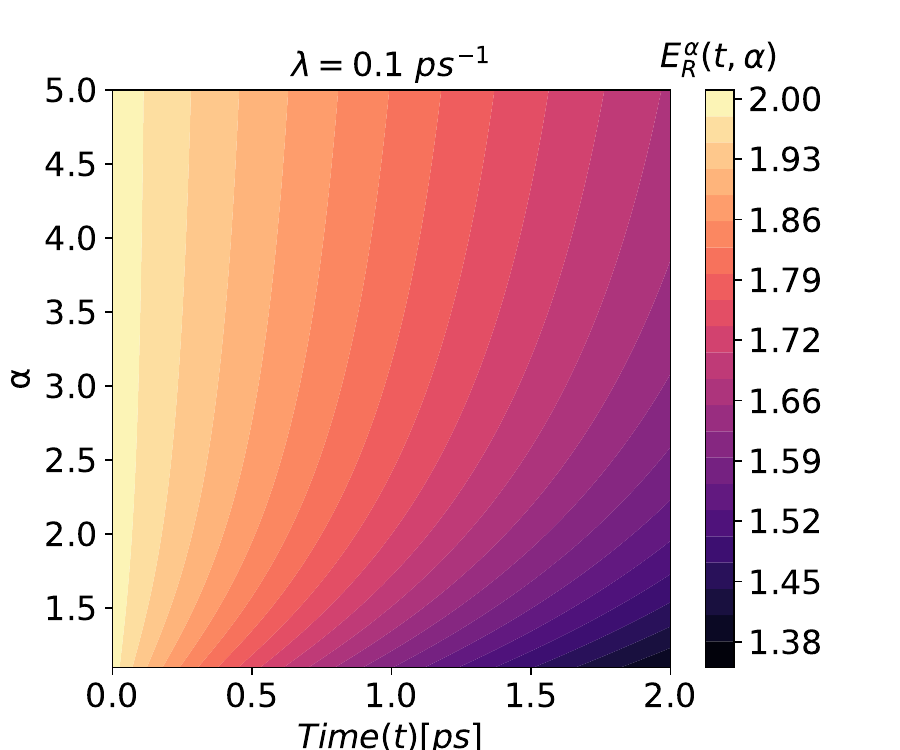}    
    \includegraphics[width=0.32\textwidth]{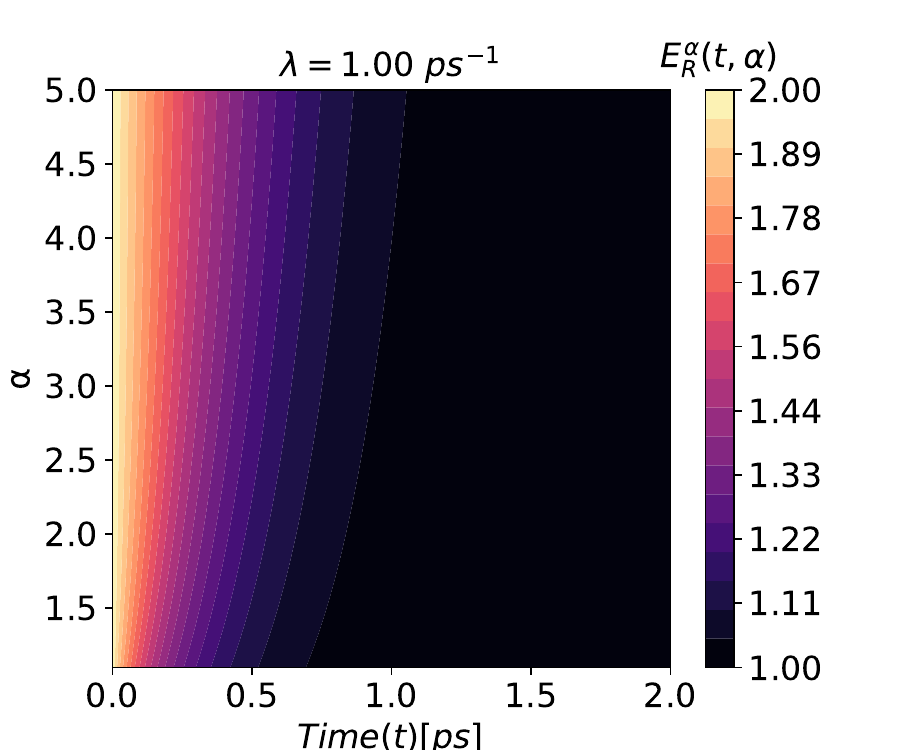}
     \caption{Relative Rényi entanglement entropy as a function of the decay time $t$ and the Rényi order $\alpha$ for different values of the decoherence parameter $\lambda$.}
    \label{fig:relative_renyi_entropy}
\end{figure*}

\subsection{Entanglement Entropy}
A bipartite system is described by a Hilbert space that factorizes as $\mathcal{H}_{AB}=\mathcal{H}_A\otimes\mathcal{H}_B$. For a general (pure or mixed) state $\rho$ of the entire system, the state of a subsystem is obtained using the partial trace, e.g., $\rho_A = \mathrm{Tr}_B(\rho)$. For pure bipartite states, a standard measure of quantum correlations is the entropy of entanglement, defined as the von Neumann entropy of the reduced state. Writing a pure state in its Schmidt form,
\begin{equation}
|\psi\rangle = \sum_i \sqrt{s_i}\,|a_i\rangle\otimes|b_i\rangle,
\end{equation} 
with Schmidt eigenvalues $s_i$ satisfying $\sum_i s_i = 1$, the reduced density matrix is $\rho_A=\sum_i s_i|a_i\rangle\langle a_i|$, and the entropy of entanglement is
\begin{equation}
S_E = S(\rho_A) = -\sum_i s_i \log_2 s_i.
\end{equation}

Its value ranges from $0$ for separable states to $\log d$ for maximally entangled states, where $d$ denotes the dimension of the smaller subsystem. At $t=0$, the $B$-meson pair is prepared in a pure state, for which the von Neumann entropy of the reduced state is a bona fide entanglement measure~\cite{Fabbrichesi:2023idl,Gabrielli:2024kbz}, a non-zero value unambiguously signals quantum correlations between the two mesons. This validity, however, is specific to the pure-state limit. Applying this measure to the entangled \BzBzbar\ system described by $\rho_E(t)$ in Eq.~\eqref{denm-sim}, the reduced state for particle $A$ is obtained from $\rho_A(t)=\mathrm{Tr}_B[\rho_E(t)]$. Because of the symmetry of $\rho_E(t)$, the reduced state takes the maximally mixed form $\rho_A(t)=\frac{1}{2} I_2$, independent of time. Its eigenvalues are therefore both equal to $1/2$, yielding $S(\rho_A)= 1$, which corresponds to the maximum value of von Neumann entropy for a two-dimensional system.

For an initially pure state $\rho_{AB}(0)$, this corresponds to one ebit of entanglement and reflects the maximally entangled nature of the \BzBzbar\ pair, analogous to a Bell state. However, once decoherence renders the global two-particle state $\rho_{E}(t)$ mixed, the reduced density matrix remains maximally mixed for all time $t$ due to the symmetry of the evolution. Since the von Neumann entropy depends only on the eigenvalues of $\rho_A(t)$, which are time-independent, it cannot register the effects of decoherence. In this regime, decoherence is present in the global state $\rho_{E}(t)$ through the damping of off-diagonal terms, but it is not captured at the single-particle level by the reduced-state entropy. 

This highlights a limitation of local entropy-based measures in quantifying entanglement in the decohering $B$-meson system. While the actual loss of entanglement is expected in the two-particle system under the open quantum system consideration, this von Neumann entropy is unable to capture the degradation of quantum correlations and instead only reflects the information about the subsystem. This motivates the use of entanglement metrics that are sensitive to the global mixed state itself, discussed in the following subsections. 

\subsection{Rényi Entropy of Entanglement}
The Rényi entropy of entanglement is a generalization of the von Neumann entropy that provides a family of entanglement measures parameterized by a real number
$\alpha \neq 1$ as
\begin{equation}
S_{\alpha}(\rho_A)=\frac{1}{1-\alpha}\log_2\!\left[\mathrm{Tr}(\rho_A^{\alpha})\right],
\end{equation}
where $\rho_A=\mathrm{Tr}_B(\rho)$ is the reduced state of subsystem $A$. In the limit
$\alpha \rightarrow 1$, $S_{\alpha}$ reduces to the von Neumann entropy. For $\alpha = 2$,
the Rényi entropy takes the form $S_2(\rho_A) = -\log \mathrm{Tr}(\rho_A^2)$. For a maximally
entangled two-level pure state, the reduced density matrix is
$\rho_A = \tfrac{1}{2} I_2$, and one finds
\[
\mathrm{Tr}(\rho_A^{\alpha}) = 2^{1-\alpha},
\]
which yields
\begin{equation}
S_{\alpha}(\rho_A) = \log_2 2 ,
\end{equation}
independent of the Rényi order $\alpha$. The same result holds for the entangled \BzBzbar\ state considered here. Its reduced state likewise remains $\rho_A(t) = \tfrac{1}{2} I_2$, and therefore $S_{\alpha}(\rho_A(t)) = \log_2 2$, for all $\alpha$ and for all times. For an initially pure state, this value corresponds to one ebit of entanglement. However, once decoherence renders the global state mixed, the Rényi entropies quantify only the local mixedness of the reduced state rather than the amount of entanglement. As in the von Neumann case, the Rényi entropies are insensitive to decoherence because $\rho_A(t)$ remains maximally mixed; consequently, information about the decay of quantum correlations is contained in the global state $\rho_{E}(t)$ and is not accessible at the single-particle reduced state alone. As in the case of the von Neumann entropy discussed above, this insensitivity to $\lambda$ in the \BzBzbar\ meson system reflects the pure-state origin of the measure rather than a genuine absence of decoherence, and the same considerations regarding its limited applicability to the decohering $B$-meson state apply here.

\subsection{Relative Rényi Entropy of Entanglement}
The relative Rényi entropy of entanglement extends the relative entropy of entanglement by replacing the von Neumann entropy with the Rényi entropy. For a bipartite state $\rho_{AB}$ it is defined as;
\begin{equation}
E_{\mathrm{R}}^{\alpha}(\rho_{AB}) =
\min_{\sigma_{AB}\in\mathrm{SEP}}
\frac{1}{\alpha-1}
\log_2\!\left[\mathrm{Tr}\!\left(\rho_{AB}^{\alpha}\sigma_{AB}^{1-\alpha}\right)\right] ,
\end{equation}
which reduces to the relative entropy of entanglement in the limit $\alpha\to1$. Here, $\sigma_{AB}$ represents a separable bipartite density matrix, and SEP denotes a set of all separable bipartite density matrices on joint Hilbert space $\mathcal{H}_A\otimes\mathcal{H}_B$. This separable state $\sigma_{AB}$ belongs to SEP iff 
\begin{equation}
\sigma_{AB}
=
\sum_i p_i\,\sigma_A^{(i)}\otimes\sigma_B^{(i)},
\end{equation}
where $p_i\geq0$ for all $i$ and 
$\sum_i p_i=1$. Also, $\sigma_A^{(i)}$ and $\sigma_B^{(i)}$ are the individual density matrices of the subsystems. If we choose $\sigma_E=I/4$, the maximally mixed two-qubit state, as the reference
separable state, then the Bell state gives the reference value $E_{\mathrm{R}}^{\alpha}(\rho_{AB}(0)\|\sigma_E)=2\log_2 2$. This value serves only as a benchmark for comparing the effect of decoherence on the R\'enyi relative entropy
evaluated for the system under consideration. With this choice of reference separable state, the relative Rényi entropy for the entangled \BzBzbar\ meson system can be written as,
\begin{equation}
E_{\mathrm{R}}^{\alpha}(\rho_E (t)\|\sigma_E)
= \frac{1}{\alpha-1}
\log_2\!\left[2^{\alpha-2}\left(a_-^{\alpha}+a_+^{\alpha}\right)\right].
\end{equation}

It should be noted that the quantity calculated above follows from the specific
choice of the reference separable state $\sigma_E=I/4$ and does not correspond to the fully minimized quantity over all separable states. In the absence of decoherence, this yields $E_{\mathrm{R}}^{\alpha}(\rho_E (t)\|\sigma_E)=\log_2 4=2$, while finite decoherence leads to a monotonic decay of quantum correlations, with larger $\alpha$ displaying a slower reduction.  

From Fig.~\ref{fig:relative_renyi_entropy}, it is evident that the relative Rényi entropy of entanglement decays monotonically with time for all Rényi orders $\alpha$ once decoherence is present. An increase in the value of the decoherence parameter $\lambda$ leads to a progressively faster suppression of entanglement, indicating an accelerated loss of quantum correlations in the $B_d^0-\bar{B}_d^0$ system. For a fixed decoherence strength, larger values of $\alpha$ exhibit a slower decay, reflecting the reduced sensitivity of higher-order Rényi measures to weak residual correlations. In the strong-decoherence regime ($\lambda \sim 1~\text{ps}^{-1}$), the entanglement rapidly approaches its minimum value, signaling the effective suppression of quantum correlations in the open quantum system dynamics.

\subsection{Relative entropy of entanglement}
The relative entropy of entanglement quantifies entanglement as the minimal distinguishability between a given quantum state and the set of separable states. In contrast to reduced-state entropies and their Rényi generalizations, which depend only on local spectra, this measure captures genuinely nonclassical correlations present in mixed states through a distance in state space. The relative entropy of entanglement $\varepsilon_R(\rho)$ for a bipartite state $\rho$ is defined as
\begin{equation}
\varepsilon_R(\rho) = \min_{\sigma \in \mathrm{SEP}} S(\rho \| \sigma),
\end{equation}
where the minimization is taken over all separable states and $S(\rho \| \sigma) = \mathrm{Tr}\!\left( \rho \log \rho - \rho \log \sigma \right)$ is the quantum relative entropy. If we choose $\sigma = I/4$, the maximally mixed two-qubit state, as the reference separable state, then the Bell state has eigenvalues $(1,0,0,0)$, while $\sigma$ has eigenvalues $(\tfrac{1}{4},\tfrac{1}{4},\tfrac{1}{4},\tfrac{1}{4})$. Substituting these into the relative entropy yields $\varepsilon_R(\rho_{AB}(0)\|\sigma) = \log_2 4 = 2$. This value serves only as a benchmark for comparing the effect of decoherence on the relative entropy evaluated for the system under consideration. Unlike the von Neumann and Rényi entropies of reduced states, which depend only on local mixedness, the relative entropy of entanglement remains sensitive to decoherence because it depends explicitly on the global two-particle state. As a result, it provides a more faithful quantifier of entanglement degradation in open quantum systems. For the entangled \BzBzbar\ meson system, the spectrum of the evolved density matrix $\rho_E(t)$ is given by
\[
\lambda_1 = \frac{1 + e^{-2\lambda t}}{2}, \qquad
\lambda_2 = \frac{1 - e^{-2\lambda t}}{2}, \qquad
\lambda_3 = \lambda_4 = 0 .
\]
while the reference separable state $\sigma_E = \tfrac{1}{4}I$ has four degenerate eigenvalues equal to $\tfrac{1}{4}$. Using the definition
\[
\varepsilon_R(\rho_E(t)) = \mathrm{Tr}\!\left[\rho_E(t)\log_2\rho_E(t)
- \rho_E(t)\log_2\sigma_E\right],
\]
one obtains
\begin{eqnarray}
\varepsilon_R(\rho_E(t)) &=&
\frac{1 + e^{-2\lambda t}}{2}
\log_2\!\left(\frac{1 + e^{-2\lambda t}}{2}\right)\nonumber\\
&+&
\frac{1 - e^{-2\lambda t}}{2}
\log_2\!\left(\frac{1 - e^{-2\lambda t}}{2}\right)
+ 2.
\end{eqnarray}

In the absence of decoherence $(\lambda = 0)$, the expression reduces to $\varepsilon_R = 2$, thereby recovering the maximal entanglement characteristic of the Bell state. For $\lambda \neq 0$, the presence of decoherence modifies the global eigenvalue spectrum, resulting in a monotonic decrease of the relative entropy of entanglement. In the long-time limit, one finds that $\varepsilon_R(\rho_E(t)) \to 1$, which signifies the progressive degradation of quantum correlations and the approach of the state to a separable regime. The temporal evolution of the relative entropy is depicted in Fig.~\ref{fig:formation_tangle}(a).

\subsection{Entanglement of Formation, Entanglement Cost, and Concurrence}
Entanglement of formation quantifies the minimal amount of entanglement required to
prepare a given bipartite quantum state using LOCC. For a mixed state $\rho_{AB}$, it is defined through the convex-roof
construction

\begin{equation}
E_F(\rho_{AB}) =
\inf_{\{p_i,|\psi_i\rangle\}}
\sum_i p_i\,
S\!\left(\mathrm{Tr}_B |\psi_i\rangle\langle\psi_i|\right),
\end{equation}

where the infimum is taken over all pure-state decompositions
$\rho_{AB}=\sum_i p_i |\psi_i\rangle\langle\psi_i|$ and
$S(\rho)=-\mathrm{Tr}(\rho\log_2\rho)$ denotes the von Neumann entropy.

\begin{figure*}[htb]
\centering

\includegraphics[width=0.4\textwidth]{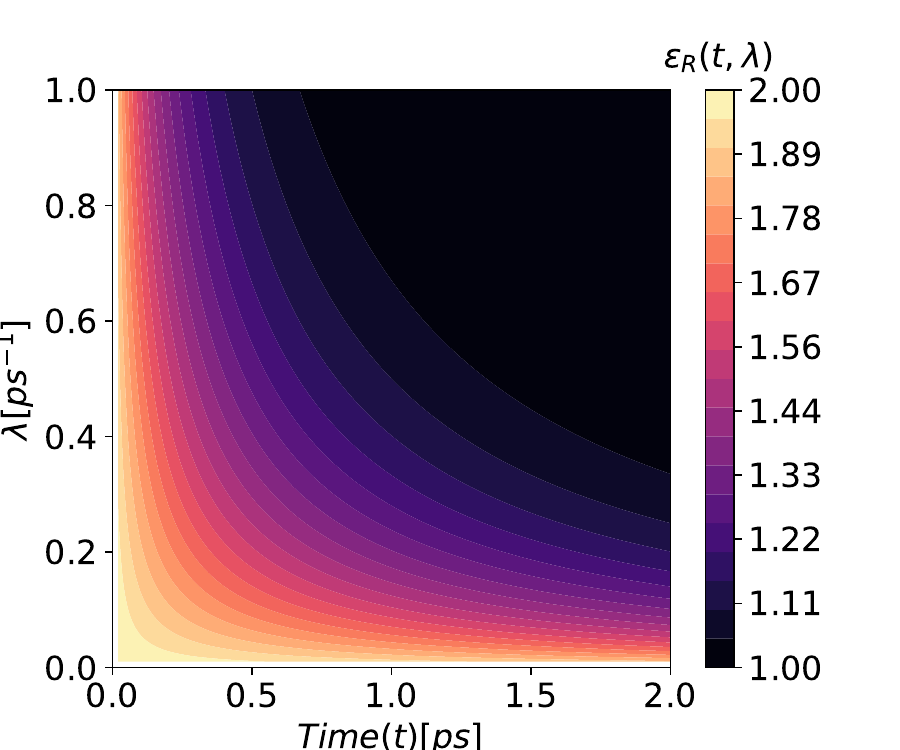}
\includegraphics[width=0.4\textwidth]{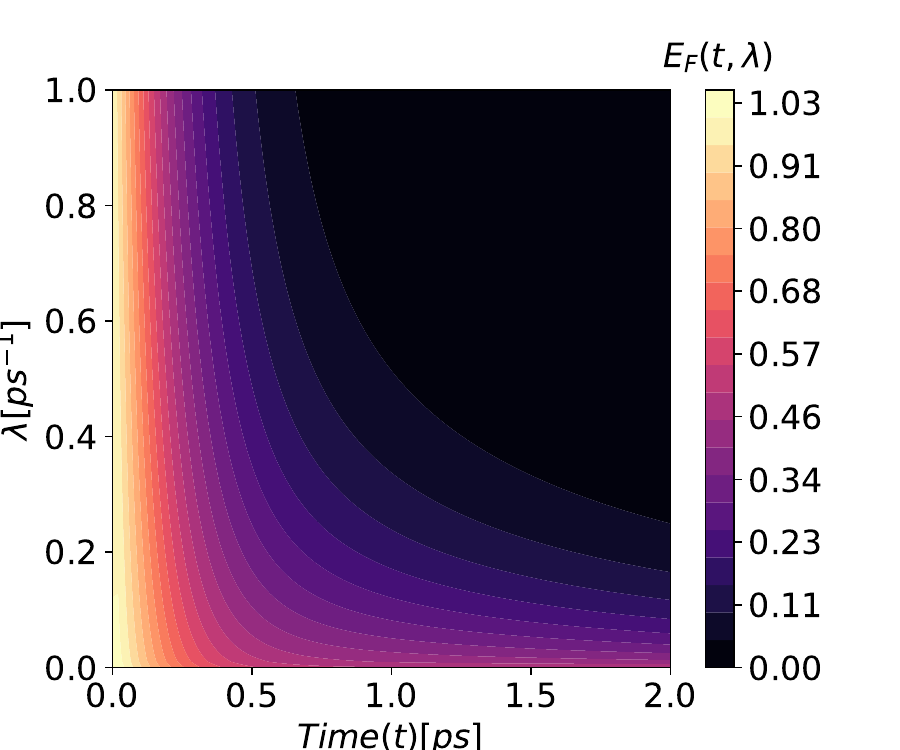}
\vspace{0.3cm}
\includegraphics[width=0.4\textwidth]{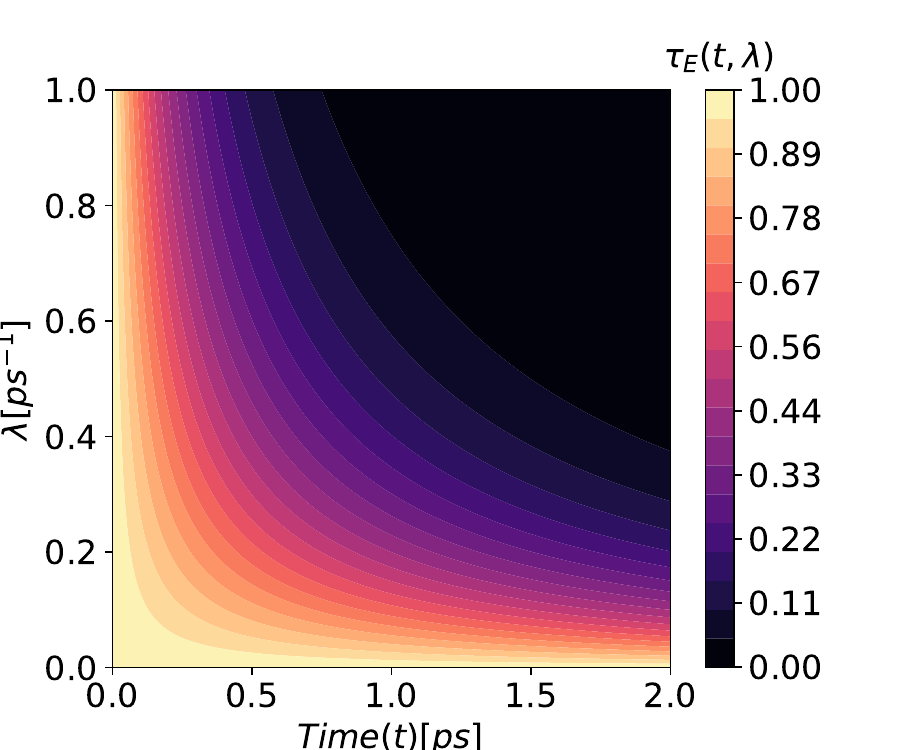}
\includegraphics[width=0.4\textwidth]{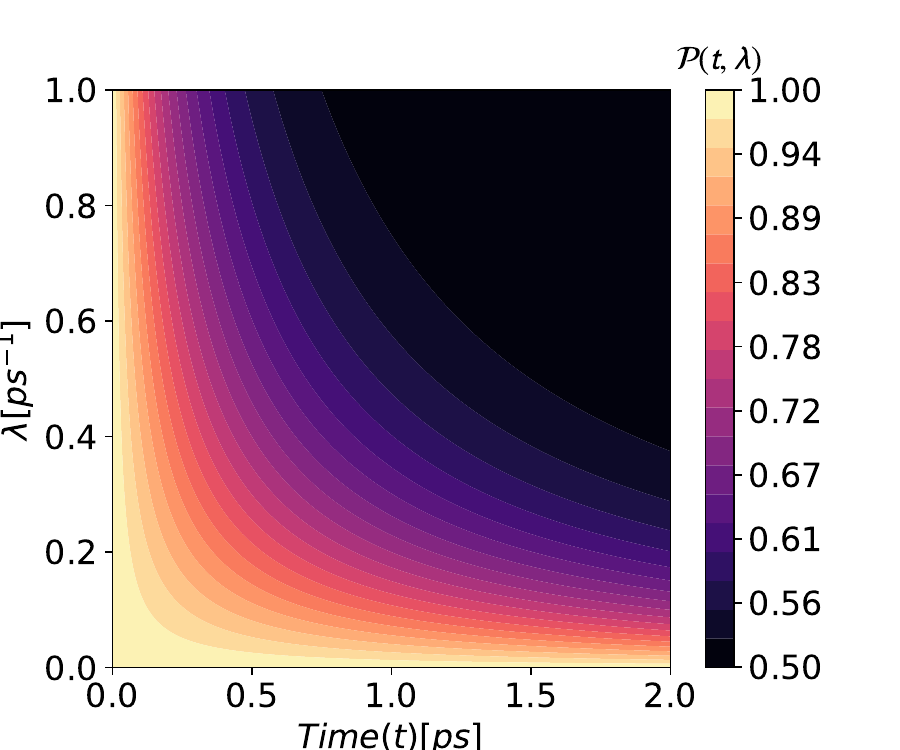}

\caption{
Contour plots of entanglement measures for the $B_d^0-\bar{B}_d^0$ system as functions of decay time $t$ and decoherence parameter $\lambda$: (a) Relative entropy of entanglement $\varepsilon_R$, (b) Entanglement of formation $E_F$, (c) $\pi$-tangle $\tau_E$,  and (d)  Purity $\mathcal{P}$.The results illustrate the degradation of quantum correlations with increasing decoherence.
}
\label{fig:formation_tangle}

\end{figure*}

Although this variational problem is generally difficult, a closed-form expression exists for two-qubit states in terms of the concurrence, which serves as an equivalent
entanglement monotone in this restricted setting. The concurrence of a two-qubit state $\rho_{AB}$ is defined as
\begin{equation}
C(\rho_{AB}) =
\max\!\left(0,\lambda_1-\lambda_2-\lambda_3-\lambda_4\right),
\label{concurr}
\end{equation}

where $\lambda_i$ are the square roots of the eigenvalues of $R=\rho_{AB}(\sigma_y\otimes\sigma_y)\rho_{AB}^*(\sigma_y\otimes\sigma_y)$ in decreasing order. Several generalizations of concurrence, such as $K$-, $I$-, and $G$-concurrence,
have been proposed for higher-dimensional or multipartite systems; however, in the present two-qubit setting, they do not provide information beyond the standard concurrence and are therefore not considered further.

In terms of the concurrence, the entanglement of formation reads
\begin{equation}
E_F(\rho_{AB}) =
h\!\left(\frac{1+\sqrt{1-C(\rho_{AB})^2}}{2}\right),
\end{equation}
with $h(x)=-x\log_2 x-(1-x)\log_2(1-x)$.

For the entangled \BzBzbar\ meson system described by $\rho_E(t)$, the concurrence
takes the simple form
\begin{equation}
C(\rho_E(t)) = e^{-2\lambda t},
\label{concur}
\end{equation}
from which the entanglement of formation follows directly. At $t=0$, one has $C=1$ and
$E_F=1$, corresponding to a maximally entangled Bell-like state. As time progresses,
environment-induced decoherence exponentially suppresses the concurrence, leading to a
monotonic decay of $E_F(\rho_E(t))$ and its eventual vanishing in the long-time limit.
This behavior is illustrated in Fig.~\ref{fig:formation_tangle}(b). Closely related to the entanglement of formation is the entanglement cost, which measures the asymptotic rate at which maximally entangled Bell pairs are required to prepare a given quantum state using LOCC. 

In general, the entanglement cost $E_C(\rho)$ is given by
the regularized entanglement of formation,
\begin{equation}
E_C(\rho) = E_F^\infty(\rho)
= \lim_{n\to\infty} \frac{1}{n} E_F(\rho^{\otimes n}),
\end{equation}
a quantity that is notoriously difficult to evaluate. For two-qubit states, however, the entanglement of formation is additive, implying
\begin{equation}
E_C(\rho)=E_F(\rho).
\end{equation}
Consequently, the results obtained above for $E_F(\rho_E(t))$ also characterize the entanglement cost of the \BzBzbar\ meson system. The decay of entanglement under decoherence, therefore, directly reflects the diminishing operational resources required to prepare the state in the open-system evolution.

\subsection{Tangle (or $\pi$-tangle)}
The tangle is an operational entanglement measure closely related to concurrence and is
particularly useful for quantifying entanglement in bipartite and multipartite systems. For two-qubit states, the tangle is defined as the square of the concurrence,
\begin{equation} 
\tau(\rho_{AB}) = C(\rho_{AB})^2,
\end{equation} 
where $C(\rho_{AB})$ is given by Eq.~\eqref{concurr}. In this setting, the tangle provides a direct measure of the strength of entanglement and is nonzero if and only if the state is entangled. For a maximally entangled Bell state, the concurrence is $C=1$, yielding a maximal tangle
$\tau=1$. For the entangled \BzBzbar\ meson system described by $\rho_E(t)$, the concurrence is derived as Eq.~\eqref{concur} which leads directly to the tangle
\begin{equation}
\tau_E(t) = e^{-4\lambda t}.
\end{equation}

At $t=0$, the tangle attains its maximal value $\tau_E=1$, reflecting the Bell-like nature
of the initial state. As time increases, decoherence exponentially suppresses the tangle, signaling the progressive loss of bipartite entanglement. This behavior is illustrated in the Fig.~\ref{fig:formation_tangle}(c), where the decay rate is governed by the decoherence parameter $\lambda$.

\subsection{Purity and Mixedness Dynamics}

Although purity is not an entanglement measure, it provides a useful diagnostic of decoherence-induced mixedness and allows one to disentangle the loss of coherence from the loss of entanglement. The purity of the
two-meson state is defined as
\begin{equation}
\mathcal{P}(t)=\mathrm{Tr}\!\left[\rho_E(t)^2\right],
\end{equation}
which equals unity for pure states and decreases as the state becomes
mixed. Using the density matrix $\rho_E(t)$ of Eq.~\eqref{denm-sim}, one
finds
\begin{equation}
\mathcal{P}(t)=\frac{1}{2}\left(1+e^{-4\lambda t}\right),
\end{equation}
while the concurrence evolves as \(C(\rho_E(t))=e^{-2\lambda t}.\)

At $t=0$ the state is pure and maximally entangled, with
$\mathcal{P}(0)=1$ and $C(0)=1$. As time progresses, decoherence drives
the system toward a mixed state with $\mathcal{P}\to 1/2$, while
entanglement vanishes only asymptotically. Fig.~\ref{fig:formation_tangle}(d) illustrates the time evolution of the purity as a function of the decoherence strength $\lambda$. For fixed $\lambda$, $\mathcal{P}(t)$ decreases monotonically from unity toward its asymptotic
value $1/2$, signaling the progressive loss of coherence. Increasing $\lambda$ accelerates this transition, confirming that mixedness growth is directly driven by environmental decoherence. In comparison with the slower decay of the concurrence, the figure makes explicit that the onset
of mixedness precedes the complete loss of entanglement.

\subsection{Schmidt Coefficients}
Schmidt coefficients provide a complete characterization of entanglement for pure
bipartite quantum states. They represent structural descriptors of bipartite correlations and serve as the spectral building blocks from which entropy-based entanglement measures are constructed, rather than constituting an operational or distance-based measure themselves. According to the Schmidt decomposition, any pure state

$|\psi\rangle \in \mathcal{H}_A \otimes \mathcal{H}_B$ can be written as
\begin{equation}
|\psi\rangle = \sum_i s_i\, |a_i\rangle \otimes |b_i\rangle,
\end{equation}
where $\{|a_i\rangle\}$ and $\{|b_i\rangle\}$ are orthonormal bases of the two subsystems and $s_i \ge 0$ are the Schmidt coefficients satisfying $\sum_i s_i^2 = 1$. The number of nonzero coefficients defines the Schmidt rank and determines whether the state is separable or entangled. For a maximally entangled two-level pure state, such as a Bell state, the Schmidt coefficients are equal, $s_1 = s_2 = 1/\sqrt{2}$. The same coefficients characterize the initial \BzBzbar\ meson pair at production, reflecting maximal entanglement at the pure-state level.

In the presence of decoherence, the global \BzBzbar\ state becomes mixed and the Schmidt
decomposition no longer applies as an entanglement measure. Nevertheless, the reduced
single-particle state remains maximally mixed, with eigenvalues $(1/2,1/2)$, leading to
formal Schmidt coefficients identical to the Bell-state values. This constancy reflects
the symmetry of the reduced dynamics rather than the persistence of entanglement, and
highlights the limitation of Schmidt coefficients in open quantum systems.

\subsection{Distance-Based Entanglement Measures}
Distance-based entanglement measures quantify entanglement in geometric terms by evaluating how far a given quantum state lies from the set of separable states under a chosen metric. In contrast to entropy-based measures, these quantities provide a direct
geometric interpretation of entanglement and can exhibit distinct sensitivities to
decoherence effects. For a quantum state $\rho$, a generic distance-based measure is defined as
\begin{equation}
E_G(\rho) = \min_{\sigma \in \mathrm{SEP}} D(\rho,\sigma),
\end{equation}
where $D(\rho,\sigma)$ denotes a distance metric and $\sigma$ is a separable reference state. In this work, we consider the Hilbert--Schmidt distance, the trace distance, and the Bures distance.

\medskip
\noindent\textit{Hilbert--Schmidt distance.}
The Hilbert--Schmidt distance between two density matrices $\rho$ and $\sigma$ is defined as
\begin{equation}
D_{HS}(\rho,\sigma)=\sqrt{\mathrm{Tr}\!\left[(\rho-\sigma)^2\right]}.
\end{equation}
For the Bell state $\rho$ and the maximally mixed separable state
$\sigma=\tfrac{1}{4}I$, the difference $\rho-\sigma$ takes the form
\begin{equation}
\rho-\sigma=\frac{1}{4}
\begin{pmatrix}
1 & 0 & 0 & 2 \\
0 & -1 & 0 & 0 \\
0 & 0 & -1 & 0 \\
2 & 0 & 0 & 1
\end{pmatrix}.
\end{equation}
Squaring this matrix and taking the trace yields
$\mathrm{Tr}\!\left[(\rho-\sigma)^2\right]=\frac{3}{4}$,
and hence $D_{HS}(\rho,\sigma)=\frac{\sqrt{3}}{2}$.
For the entangled meson state $\rho_E(t)$ and $\sigma_E=\tfrac{1}{4}I$, the squared difference
matrix is given by
\begin{equation}
(\rho_E(t)-\sigma_E)^2=
\left(
\frac{1}{4}
\begin{pmatrix}
a_- - 1 & 0 & 0 & -a_- \\
0 & a_+ - 1 & -a_+ & 0 \\
0 & -a_+ & a_+ - 1 & 0 \\
-a_- & 0 & 0 & a_- - 1
\end{pmatrix}
\right)^2 ,
\end{equation}
where $a_{\pm}=(1\pm e^{-2\lambda t})/2$.
Evaluating the trace gives
$\mathrm{Tr}\!\left[(\rho_E(t)-\sigma_E)^2\right]
=\frac{1}{4}\left(1+2e^{-4\lambda t}\right)$,
leading to
$D_{HS}(\rho_E(t),\sigma_E)
=\frac{1}{2}\sqrt{1+2e^{-4\lambda t}}$
This reproduces the Bell-state value $\sqrt{3}/2$ at $t=0$ and decreases
monotonically under decoherence, approaching the asymptotic value $1/2$.

\medskip

\noindent\textit{Trace distance.}
The trace distance between $\rho$ and $\sigma$ is defined as
\begin{equation}
D_T(\rho,\sigma)=\frac{1}{2}\mathrm{Tr}|\rho-\sigma|.
\end{equation}
For the Bell state, the operator $\rho-\sigma$ has eigenvalues
$\{3/4,-1/4,-1/4,-1/4\}$. The trace distance, defined as one-half of the
sum of the absolute eigenvalues, therefore reduces to
\begin{equation}
D_T(\rho,\sigma)=\frac{3}{4}.
\end{equation}
For the entangled meson state, the eigenvalues of
$\rho_E(t)-\sigma_E$ depend explicitly on the evolution time. Summing their absolute values yields
\begin{equation}
\sum_{i=1}^{4}|\lambda_i|=
\begin{cases}
e^{-2\lambda t}+\tfrac{1}{2}, &
t<\tfrac{\log 2}{2\lambda},\\[4pt]
1, &
t>\tfrac{\log 2}{2\lambda}.
\end{cases}
\end{equation}
Consequently, the trace distance takes the form
\begin{equation}
D_T(\rho_E(t),\sigma_E)=
\begin{cases}
\frac{1}{4}\!\left(1+2e^{-2\lambda t}\right), &
t<\tfrac{\log 2}{2\lambda},\\[4pt]
\frac{1}{2}, &
t>\tfrac{\log 2}{2\lambda}.
\end{cases}
\end{equation}
It decreases from the Bell-state value $3/4$ and eventually saturates,
reflecting residual mixedness even after strong decoherence.

\medskip

\noindent\textit{Bures distance.}
The Bures distance is defined in terms of the fidelity $F(\rho,\sigma)$ as
\begin{equation}
D_B(\rho,\sigma)=2\left(1-\sqrt{F(\rho,\sigma)}\right).
\end{equation}
For $\sigma=\tfrac{1}{4}I$, the fidelity simplifies to
\begin{equation}
F(\rho,\sigma)=\left(\frac{1}{2}\mathrm{Tr}\rho\right)^2=\frac{1}{4},
\end{equation}
yielding
\begin{equation}
D_B(\rho,\sigma)=1.
\end{equation}
This result holds for both the Bell state and the entangled \BzBzbar\ meson state at
$t=0$, and remains approximately constant under the present decoherence model, indicating reduced
sensitivity to the suppression of coherence compared to the
Hilbert--Schmidt and trace distances.

Overall, distance-based measures provide complementary geometric insight into
entanglement degradation in the \BzBzbar\ system. While the Hilbert--Schmidt and trace
distances clearly capture decoherence effects, the Bures distance remains largely
insensitive, highlighting the metric dependence of geometric entanglement quantifiers.

\subsection{Global Entanglement}
Global entanglement quantifies the overall degree of entanglement in a multipartite quantum system by probing the mixedness of its subsystems, rather than pairwise correlations. A commonly used definition, due to Meyer and Wallach, is based on the average purity of single-particle reduced density matrices. For an $n$-partite state $\rho$, it is given by
\begin{equation}
E_G(\rho) = 2\left(1 - \frac{1}{n}\sum_{i=1}^{n}\mathrm{Tr}(\rho_i^2)\right),
\label{global}
\end{equation}

where $\rho_i=\mathrm{Tr}_{\overline{i}}(\rho)$ denotes the reduced state of the $i$th subsystem. For a bipartite Bell state ($n=2$), each reduced density matrix is maximally mixed, $\mathrm{Tr}(\rho_i^2)=1/2$, yielding $E_G=1$, which corresponds to maximal global
entanglement. The same situation arises for the entangled \BzBzbar\ meson system: the reduced density matrices remain maximally mixed for all times, with
$\mathrm{Tr}(\rho_i^2)=1/2$, leading to
\begin{equation}
E_G(\rho_E(t)) = 1.
\end{equation}

Thus, although decoherence suppresses global quantum correlations, the global entanglement defined via local purities remains constant. This underscores a limitation of purity-based measures in open quantum systems, as they capture local mixedness but fail to fully characterize the dynamical degradation of nonlocal entanglement.

\subsection{Negativity and Logarithmic Negativity}
Negativity and logarithmic negativity provide computable entanglement
monotones for mixed states, based on the spectrum of the partially transposed density matrix. For a bipartite state $\rho$, the negativity
is defined as
\begin{equation}
N(\rho)=\sum_{\lambda_i<0}|\lambda_i|,
\end{equation}
where $\lambda_i$ are the eigenvalues of $\rho^{T_B}$, and the
logarithmic negativity as
\begin{equation}
E_N(\rho)=\log_2\!\left(1+2N(\rho)\right).
\end{equation}

For the entangled $B^0\bar B^0$ meson state $\rho_E(t)$, the partial
transpose $\rho_E^{T_B}(t)$ has eigenvalues
\begin{equation}
\lambda_{1,2}=\pm\frac{e^{-2\lambda t}}{2}, \qquad
\lambda_{3,4}=\frac{1}{2}.
\end{equation}

\begin{figure}[H]
     \centering
     \includegraphics[width=0.9\linewidth]{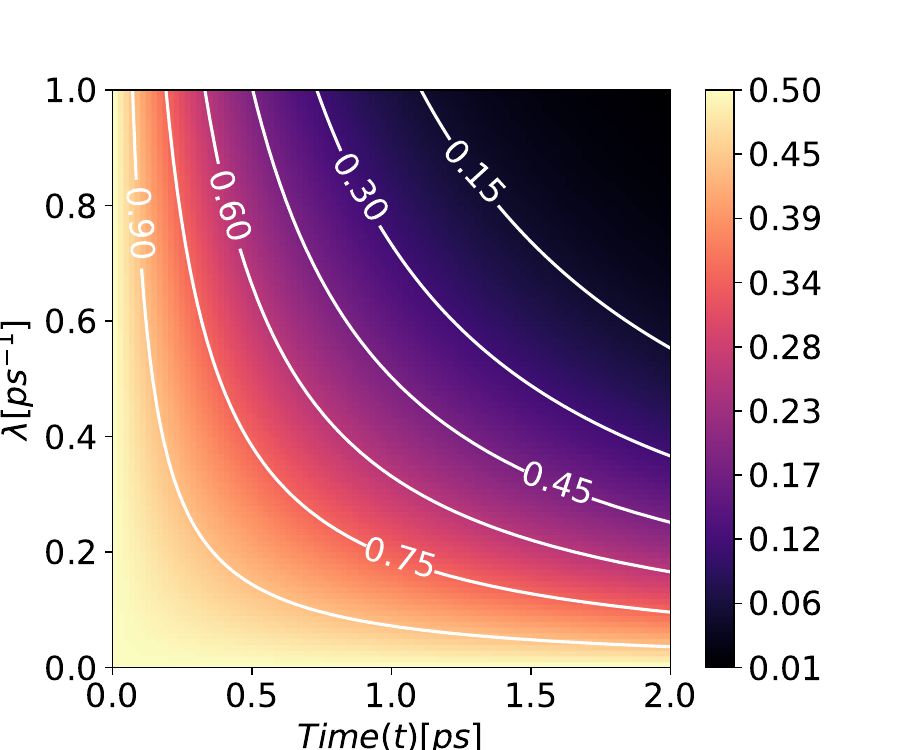}
     \caption{The heat map represents Negativity $(N)$ and the contour line in white illustrates the constant logarithmic negativity $(E_N)$ for the entangled $B_d^0-\bar{B}_d^0$  meson system,  displayed as a function of the decay time $t$ and the decoherence parameter $\lambda$.}
     \label{fig:Negativity}
 \end{figure}
 
The negativity, therefore, takes the simple form
\begin{equation}
N(\rho_E(t))=\frac{e^{-2\lambda t}}{2},
\end{equation}
from which the logarithmic negativity follows as
\begin{equation}
E_N(\rho_E(t))=\log_2\!\left(1+e^{-2\lambda t}\right).
\end{equation}

It is evident from Fig.~\ref{fig:Negativity} that both measures decay monotonically with increasing time and decoherence strength, providing a clear and time-sensitive characterization of entanglement degradation in the open-system evolution.

\section{Conclusions}
\label{sec:conc}

Neutral-meson factories furnish a unique laboratory in which both high-energy and foundational quantum-information questions can be addressed. We have presented a detailed study of quantum entanglement measures in the entangled \BzBzbar\ system, using a Kraus-operator description of the open quantum system evolution, with a decoherence parameter~$(\lambda)$ that captures environment-induced loss of coherence. For this time-dependent density matrix, we evaluated a broad selection of entanglement measures, including entropy-based measures, relative entropy of entanglement, Rényi entropies, entanglement of formation, distance-based geometric measures, negativities, purity, global-entanglement measures, and Schmidt coefficients, each capturing a different aspect of the correlations.

When $\lambda = 0$, the relevant measures reproduce the expected Bell-state behavior, confirming the maximally entangled nature of the initially produced state. For nonzero values of the decoherence parameter $\lambda$, the behavior of different quantum entanglement measures varies in a distinct manner. The entanglement entropy, Rényi entropy of entanglement, geometric entanglement (Bures distance), Schmidt coefficients, and global entanglement remain $\lambda$-independent, indicating their limitation to capture loss of entanglement in the decohering neutral meson system considered in our framework. However, the Hilbert-Schmidt and trace distance forms of the geometric measure, the relative Rényi entropy of entanglement, the entanglement of formation, the purity, the tangle, negativity, and the logarithmic negativity all exhibit explicit $\lambda$-dependence, reflecting their sensitivity to the decoherence effects in the \BzBzbar\ system.

Thus, we identify the quantum information measures which are reliable metrics of loss of entanglement due to environmental decoherence in the correlated \BzBzbar\ system. In a broad sense, the \BzBzbar\ meson system serves as an important platform for studying the interplay between entanglement and decoherence. These findings are particularly relevant in light of recent developments in high-energy quantum information.

\vspace{1em}
\section*{Acknowledgement}
This work was initiated under the guidance of Prof. Ashutosh Kumar Alok, who passed away unexpectedly during the preparation of this manuscript. We dedicate this work to his memory in recognition of his invaluable guidance and mentorship. J.K. acknowledges financial support from ANRF (formerly DST-SERB), India, under Grant No. EEQ/2023/000959, and from IIT Jodhpur, India, under Project No. I/RIG/JTK/20240067. We also acknowledge the anonymous referee, whose comments helped us improve the clarity and presentation of the manuscript.

\end{document}